\documentclass[10pt,a4paper]{article}  
\usepackage{amsmath,amssymb}
\usepackage{bm}
\usepackage{bbold}
\usepackage[margin=1.4cm]{geometry}
\usepackage{dblfnote}
\usepackage{multicol}
\usepackage[format=plain, indention=0.3cm]{caption}
\usepackage{graphicx}
\usepackage[round]{natbib}
\usepackage{paralist}
\usepackage{hyperref}

\newcommand{\bfq}{\ensuremath{\bm{q}}}
\newcommand{\bfr}{\ensuremath{\bm{r}}}
\newcommand{\bfs}{\ensuremath{\bm{s}}}
\newcommand{\bfx}{\ensuremath{\bm{x}}}

\newcommand{\bbR}{\ensuremath{\mathbb{R}}}

\DeclareSymbolFont{bbold}{U}{bbold}{m}{n}
\DeclareSymbolFontAlphabet{\mathbbold}{bbold}

\newcommand{\CW}{\ensuremath{W}}

\newcommand{\DD}{\ensuremath{\textsf{DD}}}
\newcommand{\DR}{\ensuremath{\textsf{DR}}}
\newcommand{\RR}{\ensuremath{\textsf{RR}}}
\newcommand{\DQ}{\ensuremath{\textsf{DQ}}}
\newcommand{\QQ}{\ensuremath{\textsf{QQ}}}

\newcommand{\Nr}{{\ensuremath{N_\mathrm{r}}}}
\newcommand{\Nq}{{\ensuremath{N_\mathrm{q}}}}

\newcommand{\mDelta}{\ensuremath{\delta}}

\newenvironment{Figure}
  {\par\medskip\noindent\minipage{\linewidth}}
  {\endminipage\par\medskip}

\begin{document}

\title{Increased accuracy of the
  two-point correlation function at no extra cost}

\author{Martin Kerscher \\
  Ludwig--Maximilians Universt\"at M\"unchen, \\
  Fakult\"at f\"ur Physik, Schellingstr. 4, D-80799 M\"unchen\\
  martin.kerscher@lmu.de}
\date{January 11, 2024}

\twocolumn[
\maketitle
%
\begin{center}
  \begin{minipage}{0.6\textwidth}
    Using the pair-count implementaion from the \texttt{Corrfunc}
    package we show that with a low discrepency sequence we can
    calculate the two-point correlation function more accurately than
    with random points at no extra computational cost.\\
  \end{minipage}
\end{center}
]

\section{Introduction}  

The \texttt{Corrfunc} package \citep{sinha:corrfunc,sinha:corrfunc2}
offers one of the fastest implementation of pair-counts and is
frequently used for calculating the two-point correlation function for
large-scale-structure statistics in cosmology.
We illustrate how one can combine the conceptual improvements
suggested by \citet{kerscher:improving} with the blazingly fast
pair-counts from
\texttt{Corrfunc}\footnote{\url{https://github.com/manodeep/Corrfunc}}.
With this approach we are increasing the numerical accuracy of the
pair-counts and we are reducing the systematic error in calculations
of the two-point correlation function.

The two-point correlation function of the galaxy distribution is often
used to constrain models of structure and galaxy formation and to
estimate parameters of cosmological models.  Observations give us the
positions of galaxies in space. The two-point density
\begin{equation}
  \varrho_2(\bfx_1,\bfx_2) = \varrho^2\ \left(1+\xi(|\bfx_1-\bfx_2|)\right)
\end{equation}
is the probability of finding two galaxies at $\bfx_1$ and $\bfx_2$,
where $\varrho$ is the number density and $\xi(r)$ is the two-point
correlation function. In a homogeneous and isotropic point process
$\xi(r)$ only depends on the separation $r=|\bfx_1-\bfx_2|$.
We use estimators to determine $\xi(r)$ from a galaxy catalogue within
a finite domain $\CW\subset\bbR^3$.  In cosmology estimators based on
random point sets are most commonly used. These rely on the data--data
$\DD$, data--random $\DR$, and random--random $\RR$ pair-counts.

With $\DR$ and $\RR$ one performs a Monte-Carlo integration using
random points \citep{kerscher:improving,kerscher:twopoint}.  Replacing
these random point sets with a low-discrepancy sequence of points
results in a quasi Monte-Carlo integration. This leads to an
improved scaling of the error that is almost proportional to $1/\Nq$,
where $\Nq$ is the number of points from this low-discrepancy
sequence.

In Sec.\,\ref{sec:qMC} we give the relevant definitions, in
Sec.\,\ref{sec:comparison} we illustrate the preferable properties of
the quasi Monte-Carlo integration with a numerical example, and in
Sec.\,\ref{sec:notes} we provide details on the
implementation\footnote{see
  \url{https://github.com/makerscher/corracc} for the code.} and give
some further comments.

\section{Pair-counts and estimators}
\label{sec:qMC}
Given the data points $\{\bfx_i\}$, $i=1,\ldots,N$ within the
observational window $\bfx_i\in W$ the
\begin{equation}
\label{eq:defDD}
\DD(r) = \frac{1}{N^2} \sum_{i=1}^N\sum_{j=1, j\ne i}^N\,
\mathbb{1}_{[r,r+\mDelta]}(|\bfx_i-\bfx_j|)
\end{equation}
is the normalised number of data--data pairs with a distance of
$r=|\bfx_i-\bfx_j|$ in the interval $[r,r+\mDelta]$.  The indicator
function $\mathbb{1}_A$ of the set $A$ is defined as
$\mathbb{1}_A(q)=1$ if $q\in A$ and $0$ for $q\notin A$.
Now we consider $\Nr$ random points $\{\bfr_i\}$, $\bfr_i\in W$,
$i=1,\ldots,\Nr$ within the same sample geometry $W$ as the data and
define the normalised number of data-random pairs
\begin{equation}
  \label{eq:defDR}
  \DR(r) =  \frac{1}{N \Nr}  \sum_{i=1}^N\ \sum_{j=1}^{\Nr}\,
  \mathbb{1}_{[r,r+\mDelta]}(|\bfx_i-\bfr_j|),
\end{equation}
and similarly
\begin{equation}
  \label{eq:defRR}
  \RR(r) = \frac{1}{\Nr^2}\sum_{i=1}^{\Nr} \sum_{j=1, j\ne i}^{\Nr}
  \mathbb{1}_{[r,r+\mDelta]}(|\bfr_i-\bfr_j|),
\end{equation}
the normalised number of random--random pairs.
The \citet{landy:bias} estimator is then given by
\begin{equation}
\widehat{\xi}(r) = \frac{\DD(r)-2\DR(r)+\RR(r)}{\RR(r)} .
\end{equation}

As an alternative to the random points we consider a low discrepancy
sequence of points. We use two distinct low discrepancy sequence
$\{\bfq_i\}$ and $\{\bfs_i\}$, with $\bfq_i,\bfs_i\in W$,
$i=1,\ldots,\Nq$ (see also \citealt{davila-kurban:improved}). With one
sequence $\{\bfq_i\}$ we can define $\DQ$ in full analogy to $\DR$.
To define $\QQ$ we need both sequences,
\begin{align}
  \QQ(r)
  &= \frac{1}{\Nq^2}\sum_{i=1}^{\Nq}\sum_{j=1}^{\Nq}\,
    \mathbb{1}_{[r,r+\mDelta]}(|\bfq_i-\bfs_j|).
    \label{eq:QQ}
\end{align}
$\QQ$ is the quasi Monte-Carlo integration scheme, and
correspondingly, $\RR$ is the standard Monte-Carlo integration scheme
for the same six-dimensional volume integral (see eq.\,(15) and (16)
in \citealt{kerscher:improving}).
Now we are set to define a \citet{landy:bias} type estimator using a
low discrepancy sequence instead of random points:
\begin{equation}
\widehat{\xi}_{\text{qmc}}(r) = \frac{\DD(r)-2\DQ(r)+\QQ(r)}{\QQ(r)} .
\end{equation}

\section{Comparison}
\label{sec:comparison}

For the comparison of $\widehat{\xi}$ and $\widehat{\xi}_{\text{qmc}}$
we choose the example data set \texttt{gals\_Mr19.ff} from the
\texttt{Corrfunc} distribution as our test data
set\footnote{\url{https://github.com/manodeep/Corrfunc/blob/master/theory/tests/data/gals_Mr19.ff}}. This
simulated galaxy sample is inside a rectangular box, and we can
calculate the exact reference value $\Xi(r)$ for the
\citet{landy:bias} estimator in this simple window
\citep{kerscher:improving}.
We generate $M=500$~random point sets and also 500~randomized
low-discrepancy sets and calculate $\widehat{\xi}^{(l)}$ and
$\widehat{\xi}_{\text{qmc}}^{(l)}$ for each of the $l=1,\ldots,M$ point
sets. We use the same number of points $\Nr=\Nq$ ranging from $10^4$
to $10^7$.
To quantify the deviation from the exact value $\Xi(r)$ we use 
\begin{equation}
  \sigma_{\text{qmc}}(\Nq,r)^2 = \frac{1}{M^2} \sum_{l=1}^M
  \left(\widehat{\xi}_{\text{qmc}}^{(l)}(r) - \Xi(r) \right)^2 .
\end{equation}
and similarly $\sigma(\Nr,r)^2$.

In the following figures we compare the estimated standard errors
$\sigma(\Nr,r)$ and $\sigma_{\text{qmc}}(\Nq,r)$.  We see that with
$\widehat{\xi}_{\text{qmc}}$ we gain accuracy in all the situations.
For small radii the estimated $\sigma$ can be reduced by a factor of
2-3 using a low discrepancy sequence (see
Fig.\,\ref{fig:smallscale}). Unfortunately the scaling of
$\sigma_{\text{qmc}}$ with $\Nq$ is only slightly steeper than the
scaling of $\sigma$ with $\Nr$ in the standard estimator.  Hence for
small radii the gain in accuracy is not really convincing.

As can be seen from Fig.\,\ref{fig:largescale} this changes on large
scales.  By using a low discrepancy sequence we gain accuracy up to a
factor of 10.  To obtain a factor of 10 in accuracy with random points
requires 100 times the number of random points.  The error for the
standard estimator using random points $\sigma(\Nr,r)$ scales
proportional to $1/\sqrt(\Nr)$, whereas the error
$\sigma_{\text{qmc}}(\Nq,r)$ for a low-discrepancy sequence is almost
proportional to $1/\Nq$.
The implementation of $\widehat{\xi}_{\text{qmc}}$ follows closely the
implementation of the standard \citet{landy:bias} estimator
$\widehat{\xi}$ (see also the next section). For the same number of
points we expect a similar run-time.
For a more examples and a detailed comparison of run-times see
\citet{kerscher:improving}.

\begin{Figure}
  \begin{center}
    \includegraphics[width=0.8\textwidth]{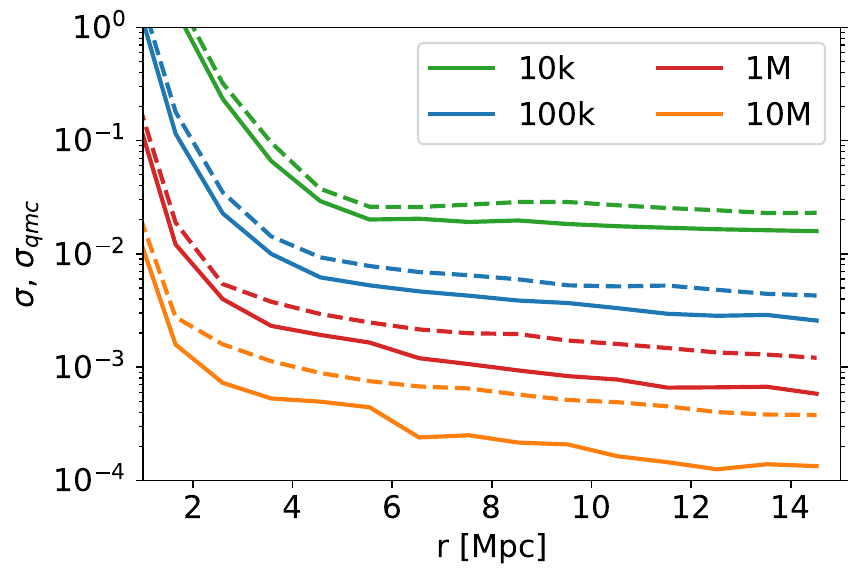}\\
    \includegraphics[width=0.8\textwidth]{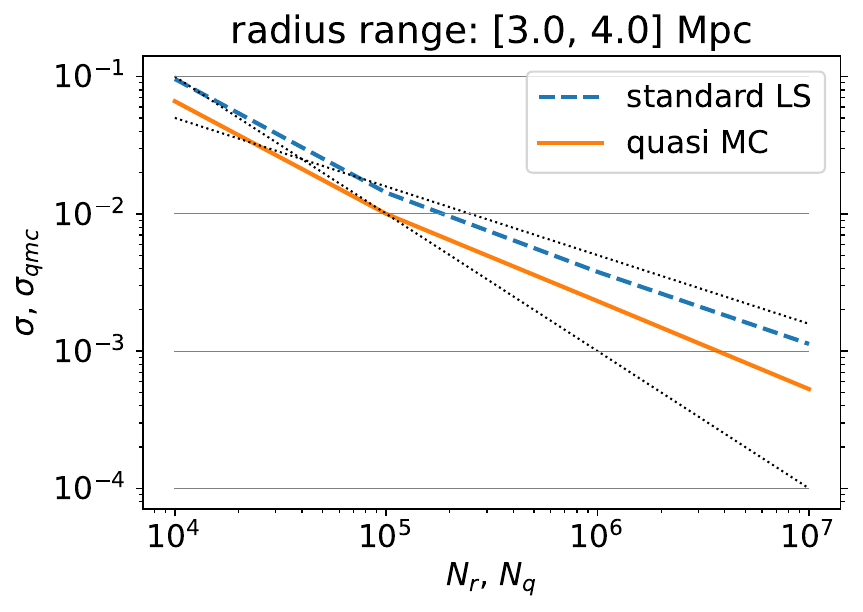}
  \end{center}
  \captionof{figure}{In the top plot we show the estimated error
    $\sigma_{\text{qmc}}(\Nq,r)$ (solid line, quasi MC) and
    $\sigma(\Nr,r)$ (dashed line, standard) for different radii
    depending on the number of points used.
    In the lower plot we show the scaling of the error with the number
    of points for one radius range.  The two dotted lines are
    proprtional to $1/\sqrt{N}$ and $1/N$.  }
  \label{fig:smallscale}
\end{Figure}

\begin{Figure}
  \begin{center}
    \includegraphics[width=0.8\textwidth]{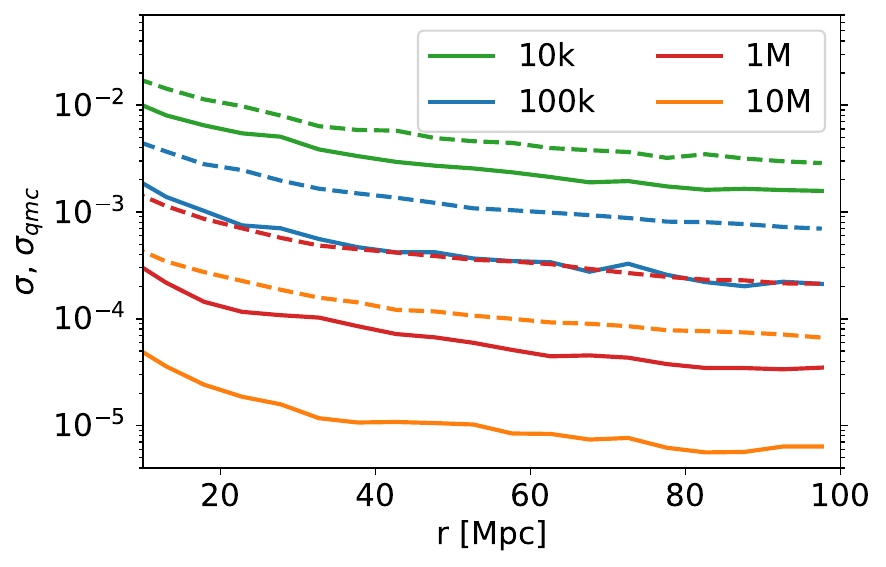}
    \includegraphics[width=0.8\textwidth]{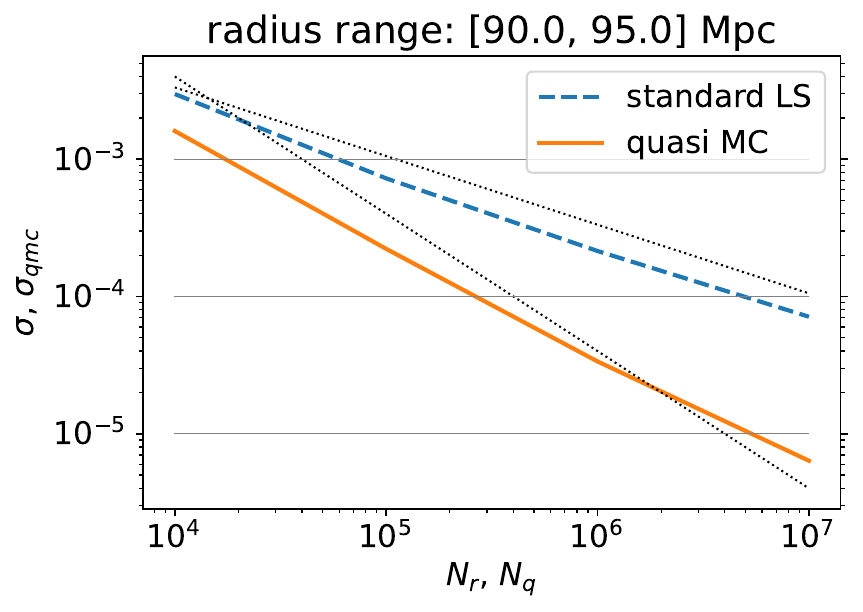}\\
  \end{center}
  \captionof{figure}{The same quantities as in
    Fig.\,\ref{fig:smallscale} are shown for larger radii.}
  \label{fig:largescale}
\end{Figure}

\section{Some notes}
\label{sec:notes}

\paragraph{The code:}
We provide a sample
implementation\footnote{\url{https://github.com/makerscher/corracc}}
to illustrate how one can use the pair-count functions as provided in
\texttt{Corrfunc} with the low discrepancy sequences from SciPy.
For calculating the auto- or cross-correlation function of point sets
the package \texttt{Corrfunc} provides the function
\texttt{Corrfunc.theory.DD}.  We use this function to calculate $\DD$,
$\RR$, $\DR$, $\DQ$ and $\QQ$.

To generate random points we use the random number
generator from NumPy \citep{harris:numpy}.
To generate low discrepancy sequences we use the function
\texttt{Halton} as provided in SciPy \texttt{scipy.stats.qmc}
(starting with version $\ge$\texttt{1.7.0},
\citealt{virtanen:scipy}). This function allows us to generate
randomised Halton sequences (see \citealt{owen:randomized},
\citealt{owen:strong}).

Special care has to be taken in the calculation of $\QQ$. As one can
see from eq.\,(\ref{eq:QQ}) we need two distinct low discrepancy
sequences (see also eq.\,(15) and (16) in
\citealt{kerscher:improving}).  We start with a six dimensional Halton
sequence inside $W\times W$ and split it into two three dimensional
sequences.

\paragraph{Exact $\Xi(r)$}
As already mentioned the $\DR$, $\RR$ and also $\DQ$, $\QQ$ are
(quasi) Monte-Carlo schemes for special volume integrals which can be
expressed in geometrical terms for a box
\citep{kerscher:improving,kerscher:twopoint}.  We use these terms to
calculate the exact estimator $\Xi(r)$ for a box.  Still a numerical
integration remains to be performed. We are using the integration
routine \texttt{quad} from \texttt{scipy.integration}. This
calculation can take up to several hours, but for the comparison we
only need this once.

\paragraph{Beyond a box}
For more general sampling areas we are not able to calculate the exact
reference value $\Xi(r)$.
Hence we have to compare to the empirical mean
\begin{equation}
  \overline{\xi}(\Nq,r)
  = \frac{1}{M} \sum_{l=1}^M \widehat{\xi}_{\text{qmc}}^{(l)}(r) ,
\end{equation}
and calculate the sample error
\begin{equation}
  \widehat{\sigma}_{\text{qmc}}(\Nq,r)^2 = \frac{1}{M^2} \sum_{l=1}^M
  \left(\widehat{\xi}_{\text{qmc}}^{(l)}(r) -
    \overline{\xi}(\Nq,r) \right)^2 .
\end{equation}
In the comparions in Sect.\,\ref{sec:comparison} we also calculated
$\widehat{\sigma}_{\text{qmc}}$ and saw that it is almost
indistinguishable from $\sigma_{\text{qmc}}$. Hence we expect that
$\widehat{\sigma}_{\text{qmc}}$ is also a good proxy for the expected
error $\sigma_{\text{qmc}}$ in more general situations. This allows
the quantification of systematic errors for a window $W$ beyond a box.

\paragraph{The periodic box:}
We compared the $\widehat{\xi}_{\text{qmc}}(r)$ and $\widehat{\xi}(r)$
for data in a box, which we saw as an example for more general windows
$W$.  Hence we ignored the fact that the data came from a simulation
with periodic boundaries.

Estimating $\xi(r)$ from data with periodic boundaries is much
simpler.  The pair-counts $\DD(r)$ have to be calculated respecting
the periodic boundary conditions which can be done also with
\texttt{Corrfunc.theory.DD}.  The geometric factors for the estimation
of $\xi(r)$ are well known and specifically simple.  If we assume a
quadratic box $[0,L]^3$ with periodic boundaries we get
\begin{align}
  \widehat{\xi}_{\text{per}}(r)
  & = \frac{L^3}{\frac{4\pi}{3} \left((r+\mDelta)^3 - r^3\right)} \DD(r)-1
\end{align}
More details are given in Appendix~A.1 of \citet{kerscher:improving}.
No random or low discrepancy point set appears, only the
pair-counts $\DD(r)$ are used, see also \texttt{Corrfunc.theory.xi}.

\paragraph{A special $\DR$ and $\DQ$}
\citet{kerscher:improving} suggests a special scheme for calculating
$\DR$ and $\DQ$ based on ideas of \citet{rivolo:two-point}.  Since
this requires a different algorithm than the pair-counts we do not
cover it here. Together with another (slower) pair-count
implementation such a special $\DR$, $\DQ$ implementation is provided
at the following
link\footnote{\url{https://homepages.physik.uni-muenchen.de/~Martin.Kerscher/software/accuratexi/}}
.


\end{document}